\documentstyle[12pt,psfig,gcdv,twocolumn]{article}
\def\reference{\parskip 0pt\par\noindent\hangindent 0.5 truecm}

\def\revised{}
\begin{document}
\small \shorttitle{Modelling Galaxies with a Multi-Phase ISM in
3d}
\shortauthor{S. Harfst et al.}
%
%
\title{Modelling Galaxies with a 3d-Multi-Phase ISM}
%


\author{\small Stefan Harfst,$^{1,2}$
 Christian Theis,$^{1,3}$ and
 Gerhard Hensler$^{1,3}$
} 

\date{}
\twocolumn[
\maketitle

{\center \footnotesize $^1$ University of Kiel, Institute of
Theoretical
Physics and Astrophysics, 24098 Kiel, Germany\\[1mm]
$^2$ Centre for Astrophysics \& Supercomputing, Swinburne
University, Mail \#31, P.O. Box 218, Hawthorn, VIC 3122,
Australia\\[1mm]
$^3$ University of Vienna, Institute of Astronomy,
T\"urkenschanzstr. 17, 1180 Vienna, Austria\\[3mm]
}

%
\begin{abstract}
We present a modified TREESPH code to model galaxies in 3d. The
model includes a multi-phase description of the interstellar
medium which combines two numerical techniques. A diffuse warm/hot
gas phase is modelled by SPH while a sticky particle scheme is
used to represent a cloudy medium. Interaction processes, such as
star formation and feedback, cooling and mixing by condensation
and evaporation, are taken into account. Here we apply our model
to the evolution of a Milky Way type galaxy. After an initial
stage, a quasi-equilibrium state is reached. It is characterised
by a star formation rate of $\sim$1\,${\rm M}_{\odot}\,{\rm
yr}^{-1}$. Condensation and evaporation rates are in balance at
0.1-1\,${\rm M}_{\odot}\,{\rm yr}^{-1}$.
\end{abstract}

\bigskip


\vspace*{1cm}

]
%
%

\section{Introduction}

In most 3d models of galaxies the interstellar medium (ISM) is
described as a single phase: either as a diffuse gas e.g.\ using
\revised{smoothed particle hydrodynamics (SPH)} (\revised{Lucy, 1977;
Gingold \& Monaghan, 1977}) or as a clumpy phase e.g.\ by sticky
particles as in Theis \& Hensler (1993). \revised{Single-phase models
have been successfully applied in the context of cosmological
simulations (Steinmetz \& M\"uller, 1994, 1995; Navarro, Frenk \&
White, 1997) as well as in simulations of isolated galaxies (Friedli
\& Benz, 1995; Raiteri, Villata \& Navarro, 1996; Berczik, 1999). A
problem in these simulation can be the so-called overcooling (White \&
Frenk, 1991; Cole et al., 1994) leading to an overproduction of dwarf
galaxies.} \revised{More recent models use a multi-phase ISM. This
allows a more realistic description of star formation (SF) and
feedback processes which could be a solution to this problem.}

\revised{A multi-phase ISM can be} accomplished by modifying the
SPH-algorithm (\revised{Hultman \& Pharasyn, 1999;} Ritchie \&
Thomas, 2001; Springel \& Hernquist, 2003). We follow a different
approach and combine both treatments of a one-phase model in a
particle based code: the hot diffuse gas phase is described by a
SPH formalism, whereas the cold molecular clouds are represented
by sticky particles. In this model, the two gas phases are
dynamical independent. The coupling between the gaseous phases is
achieved by condensation/evaporation (C/E) and by a drag force due
to ram pressure. Energy is dissipated by cloud-cloud-collisions or
by radiative cooling. Furthermore, stars are formed in clouds and
the stars return mass and energy to the gas by feedback processes
(supernovae (SNe), planetary nebulae (PNe)). A similar concept is
followed by Semelin \& Combes (2002) and Berczik et al. (2003).

\begin{figure*}[!t]
\begin{center}
\begin{tabular}{ccc}
\psfig{width=8cm,angle=0,file=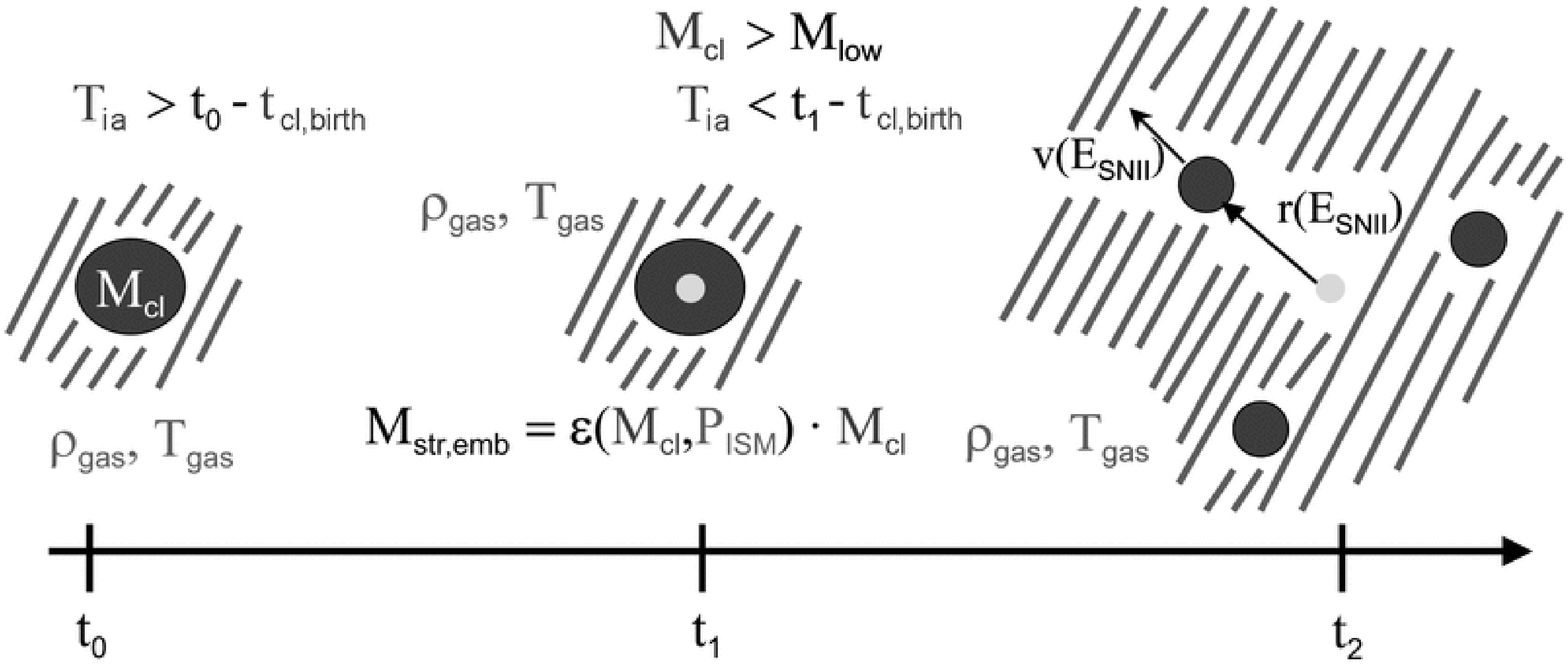} &&
\psfig{width=8cm,angle=90,file=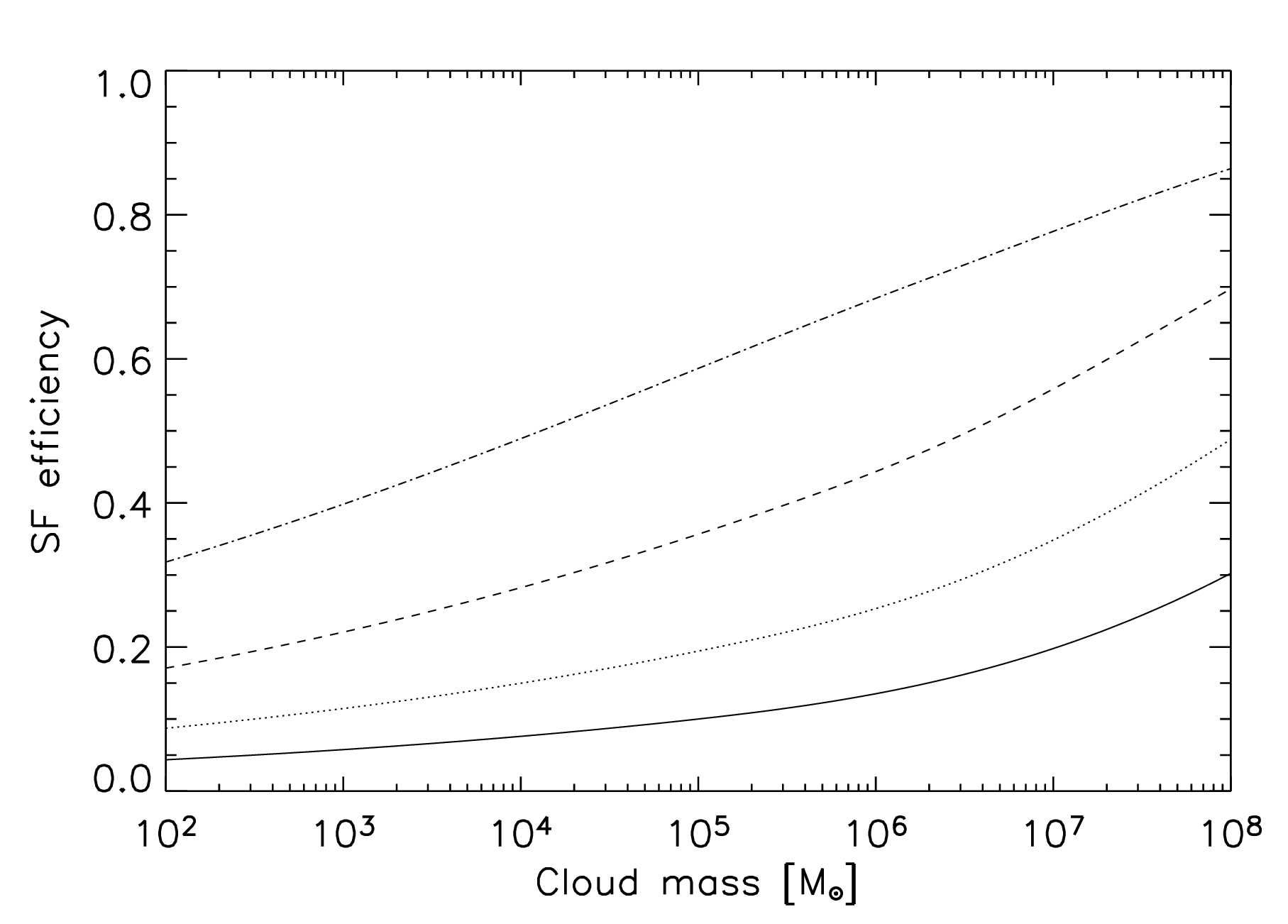} \\
\end{tabular}
\caption{ The left diagram shows the SF scheme: ($t_0$) the cloud
is inactive (no SF); ($t_1$) an embedded star cluster has been
formed with a local SF efficiency; ($t_2$) the cloud is fragmented
by SNe energy input. \revised{In the right plot the local SF
efficiency depending on the mass of the star forming cloud and the
local pressure is shown. It is based on a model proposed by
Elmegreen \& Efremov (1997).}} \label{harfst_fig1}
\end{center}
\end{figure*}

In SPH codes the star formation (SF) is usually based on the
Schmidt law, i.e.\ the SF rate depends on gas density to some
power and a characteristic time scale. For sticky particles the SF
can be coupled to cloud-cloud collisions, or a single cloud forms
stars with a constant SF efficiency. Here, a SF scheme using a
different approach is presented: the process of SF is treated
individually for each cloud. The SF efficiency depends on
local properties of the ISM and the star forming clouds, thereby
enabling self-regulation of SF by feedback.

The code is described in more detail in Sec.~\ref{harfst_sec_code}. In
Sec.~\ref{harfst_sec_res} a model of a Milky Way type galaxy is
presented. A short summary and future prospects are given in
Sec.~\ref{harfst_sec_sum}.

\section{Numerical Treatment}
\label{harfst_sec_code}

\revised{We use a particle code to model the evolution of galaxies. In
our code, different components of a galaxy are distinguished by means
of particle types. These particle types are stars, clouds, (diffuse)
gas and dark matter particles. The gravitational force for all
particles is calculated with a new, very fast TREE-algorithm proposed
by Dehnen (2002). An additional external potential can act as a static
dark matter halo. A standard SPH formalism is used to compute the
hydrodynamics of the diffuse gas. A description of the SPH method can
be found in Monaghan (1992). Furthermore, the code includes the sticky
particle scheme of Theis \& Hensler (1993) to describe the collisional
cloud dynamics.}
The integration of the system is done with a leap-frog scheme.
Individual time steps are used for each particle.

The drag force and the mass exchange rates for C/E are calculated
according to Cowie et al. (1981). We apply a mass-radius-relation for
the clouds based on observations (e.g.\ Rivolo \& Solomon,
1988). Effects of both processes are individually determined for each
cloud using the local density, temperature and velocity of the hot
gas. Energy can be dissipated by inelastic cloud-cloud collisions
(Theis \& Hensler, 1993) or by radiative cooling.  \revised{The
cooling function is taken from B\"ohringer \& Hensler (1989).}

In Fig.~\ref{harfst_fig1} the process of SF is drawn
schematically. The general idea is that stars are formed in clouds
and the cloud is fragmented in the process. Each cloud (or
fragment from a previous SF process) is inactive for a given time
${\rm T}_{\rm ia}$ (${\rm t}_0$). The time ${\rm T}_{\rm ia}$
represents a \revised{global} SF time scale \revised{which is
typically of the order of a few 100\,Myr. The time scale of the SF
process itself is neglected as recent observations suggest that SF
occurs on the much faster local crossing time scale (Elmegreen,
2000). Because not all the gas in clouds is dense or unstable
enough for SF, ${\rm T}_{\rm ia}$ is the time scale on which
clouds provide the matter that is able to form stars.}


Once the SF process has started, an embedded star cluster is formed
(${\rm t}_1$) with the SF efficiency $\epsilon$ being a function of
cloud mass and gas pressure according to Elmegreen \& Efremov
(1997). The energy released by SNe \revised{(type II)} is calculated
using a multi-power law IMF (Kroupa et al., 1993) \revised{with
stellar lifetimes taken into account (Rateri et al., 1996)}. This
energy injection disrupts the cloud into smaller fragments (${\rm
t}_2$) \revised{leaving a new star particle}. The time for disruption
${\rm T}_{\rm dis}$ (${\rm t}_2 - {\rm t}_1$ in
Fig.~\ref{harfst_fig1}) is determined from the energy input: a
self-similar solution is used to calculate the SN shell expansion.  It
is assumed that the cloud disrupts when the shell radius equals the
cloud radius. The fragments are then placed symmetrically on the shell
with velocities corresponding to the expansion velocity at ${\rm
T}_{\rm dis}$. The energy not consumed by cloud disruption and kinetic
energy of the ejecta is returned to the hot gas phase (SPH
particles). Additionally, the mass returned by PNe is added to
surrounding cloud particles \revised{also taking stellar lifetimes
into account}.


\section{The model and results}
\label{harfst_sec_res}

\begin{table}
  \centering
  \caption{Properties of the initial disk galaxy
 model}\label{harfst_tab_inimod} \setlength{\doublerulesep}{1pt}\renewcommand{\arraystretch}{1.3}
\begin{tabular}{lcc}
  \\[3pt]
  \hline\hline
  component & mass$^1$ & \#particles \\[3pt]
  disk & 0.343 & 20000 \\[-3pt]
  \hspace{5mm}- stars & 0.274 & 16000 \\[-3pt]
  \hspace{5mm}- clouds & 0.069 & 4000 \\[1pt]
  bulge & 0.167 & 10000 \\[1pt]
  DM halo & 1.94 & - \\[1pt]
  hot gas & 0.002 & 10000 \\[1pt]
  total & 0.002 & 40000 \\[3pt]
  \hline\hline
  \multicolumn{3}{l}{$^1$ in units of $10^{11}\,{\rm M}_{\odot}$} \\
\end{tabular}
\end{table}

Initially, a galaxy similar to the Milky Way was set up using the
galaxy building package described by Kuijken \& Dubinski (1995).
\revised{We chose their model A and some parameters of the initial
model are listed in Tab.~\ref{harfst_tab_inimod}.} A cloudy gas phase
is added by treating every fifth particle from the stellar disk as a
cloud. The total mass in clouds is ${\rm M}_{\rm cl,tot} \approx
6.9\cdot10^9\,{\rm M}_{\odot}$ and each cloud is randomly assigned a
time of inactivity ${\rm T}_{\rm ia}$ between 0 and 200\,Myr. Finally,
a \revised{slowly rotating} homogenous gas halo with a total mass of
${\rm M}_{\rm hot,tot} = 2\cdot10^8\,{\rm M}_{\odot}$ is
added. \revised{We do not follow the chemical evolution of the galaxy,
so where a metallicity is needed we use solar abundances. In order to
treat the feedback of 'old' star particles each star particle is
assigned an age. For the bulge stars the age is 10\,Gyr and for disk
stars 0-10\,Gyr.}

\begin{figure}[!t]
\centerline{\psfig{width=8cm,angle=90,file=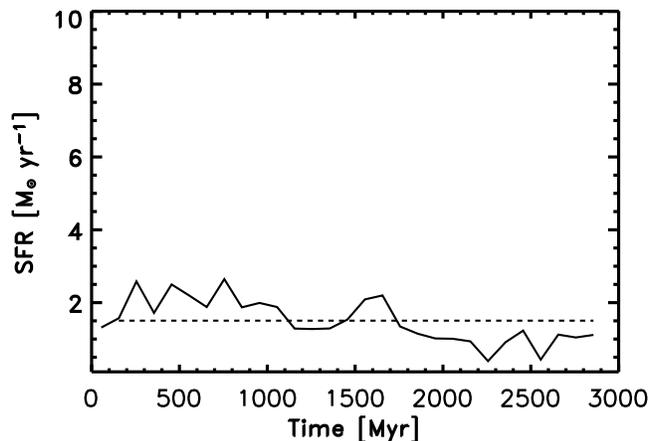}}
\caption{ Shown is the temporal evolution (solid line) and the
average (dashed line) of the SF rate for a model of Milky Way type
galaxy.} \label{harfst_fig2}
\end{figure}

Starting from this setup the galaxy model is evolved for 3\,Gyr
with all processes switched on. During an initial phase (0 to
$\sim$1\,Gyr) the gas halo collapses and expands again to reach an
equilibrium defined by the interplay of the gravitational
potential, radiative cooling, heating by SNe and
condensation/evaporation. For the following 2\,Gyr the galaxy
shows a stable evolution with an average SF rate of $\sim$1\,${\rm
M}_{\odot}\,{\rm yr}^{-1}$ (Fig.~\ref{harfst_fig2}). The SF rate
is slowly decreasing with time due to the consumption of the
clouds while the average SF efficiency is roughly constant at
$5-10$\%. \revised{In the end, the number of particles has
increased by a factor of about $2.5$. The final number of stars
and clouds are $63\,748$ and $27\,038$, respectively.}

\begin{figure*}[!t]
\begin{center}
\begin{tabular}{ccc}
\psfig{width=8cm,angle=90,file=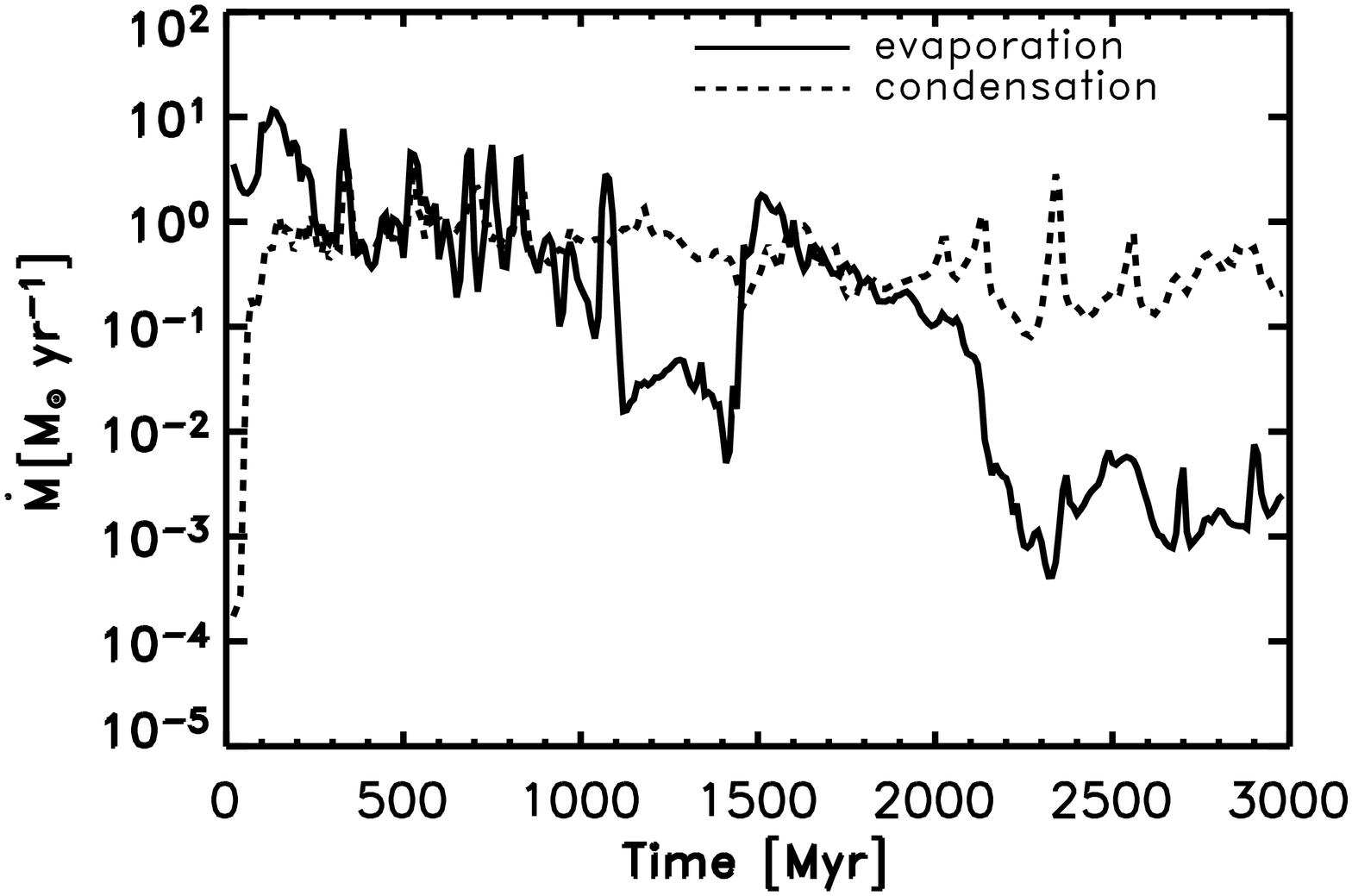} &&
\psfig{width=8cm,angle=90,file=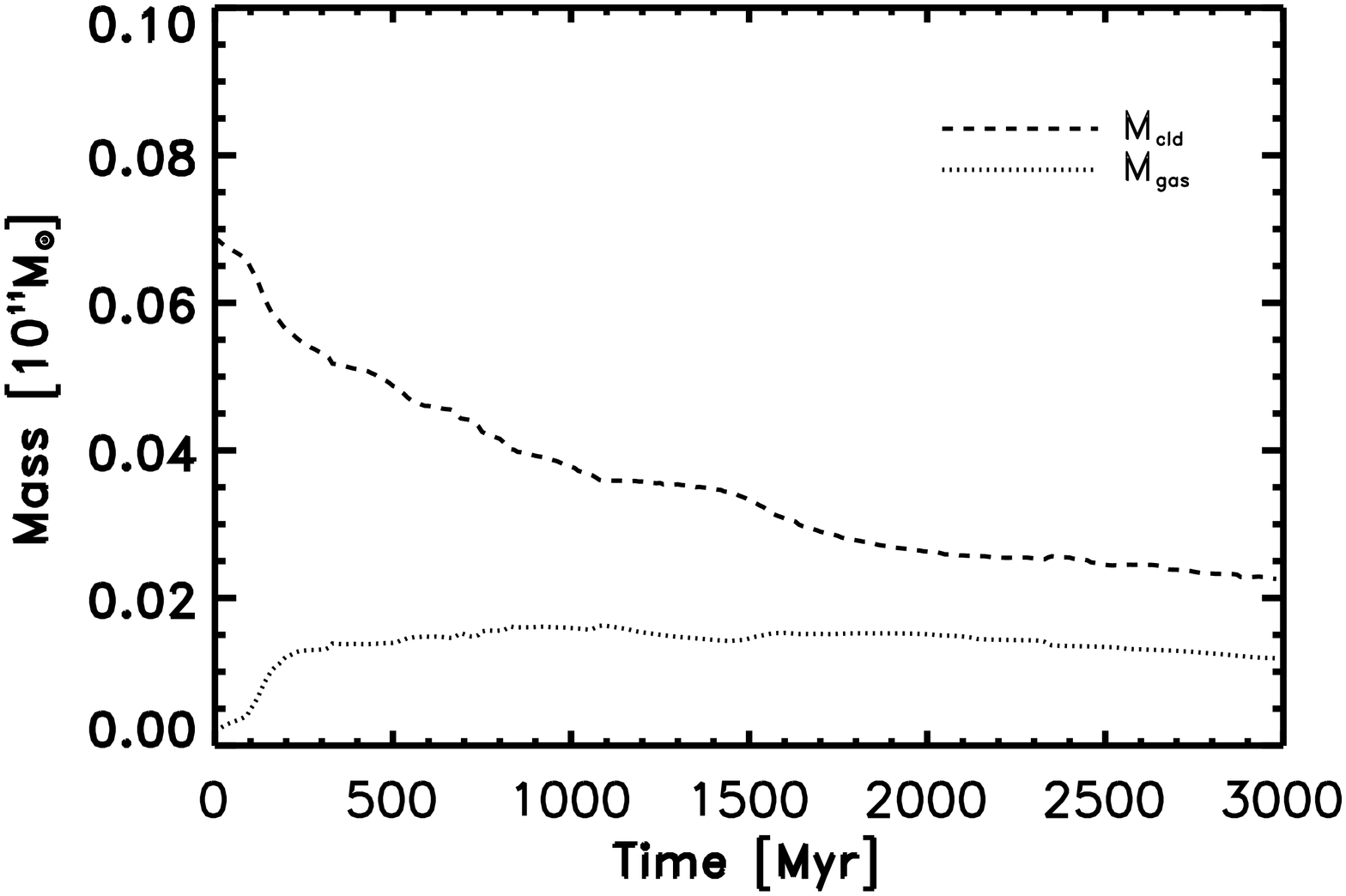} \\
\end{tabular}
\caption{ The left diagram shows the C/E rates vs. time. After an
initial phase ($<300\,{\rm Myr}$) both processes reach an equilibrium
at roughly ${1\,\rm M}_{\odot}\,{\rm yr}^{-1}$ ($300-1000\,{\rm
Myr}$). The lower evaporation rate at $\sim$1\,Gyr and $\sim$2\,Gyr
coincides with a lower SF rate. The right plot shows the total mass in
the cloudy and the diffuse ISM.  The mass of the diffuse gas is almost
constant after an early strong increase while the cloud mass is
decreasing due to SF.}
\label{harfst_fig3}
\end{center}
\end{figure*}

The mass exchange by C/E reaches a quasi-equilibrium after a few
100\,Myr (Fig.~\ref{harfst_fig3}, left panel). Before, evaporation
is the dominant process due to the initial collapse and heating of
the diffuse gas. In the later evolution, the condensation rate is
in the range of 0.1-1\,${\rm M}_{\odot}\,{\rm yr}^{-1}$ while the
evaporation rate occasionally drops by few orders of magnitude.
These drops can be correlated with a lower SF rate, i.e. a lower
heating by SNe.

The initially strong evaporation can also be seen in the comparison of
the total mass in the gas phases (Fig.~\ref{harfst_fig3}, right
panel). The diffuse gas mass increases strongly in the beginning but
is roughly constant after that. The total cloud mass is slowly
decreasing, showing the depletion by SF. A detailed look at the
diffuse gas shows also that only a small fraction (5\%) is in the
halo, while most gas (65\%) is located in the disk ($R<20\,{\rm kpc}$
and $z<1\,{\rm kpc}$). Some gas (30\%) has even left (at least
temporarily) the galaxy ($R>50\,{\rm kpc}$). The average densities in
halo and disk are $3 \cdot 10^{-4}\,{\rm cm}^{-3}$ and $1 \cdot
10^{-1}\,{\rm cm}^{-3}$, respectively. \revised{The gas in the halo is
hot ($>10^6$\,{\rm K}) while} the gas in the disk is cooler
\revised{($\sim 10^4$\,{\rm K}) and could be described as a warm
phase}. In fact, it could form clouds from thermal instabilities, so
that this process should also be taken into account in future
calculations. \revised{In the Milky Way the total amount of warm and
cold gas observed in the disk is $\sim 6\cdot10^9\,{\rm M}_{\odot}$
which is about a factor of $2$ more than what we see at the end of our
simulation. In order to get a better agreement we could increase the
initial gas mass fraction (i.e.\ select more clouds) or add an
extended gas disk as a reservoir for a continuous gas infall.}

\begin{figure*}[!t]
\begin{center}
\begin{tabular}{ccc}
\psfig{width=8cm,angle=90,file=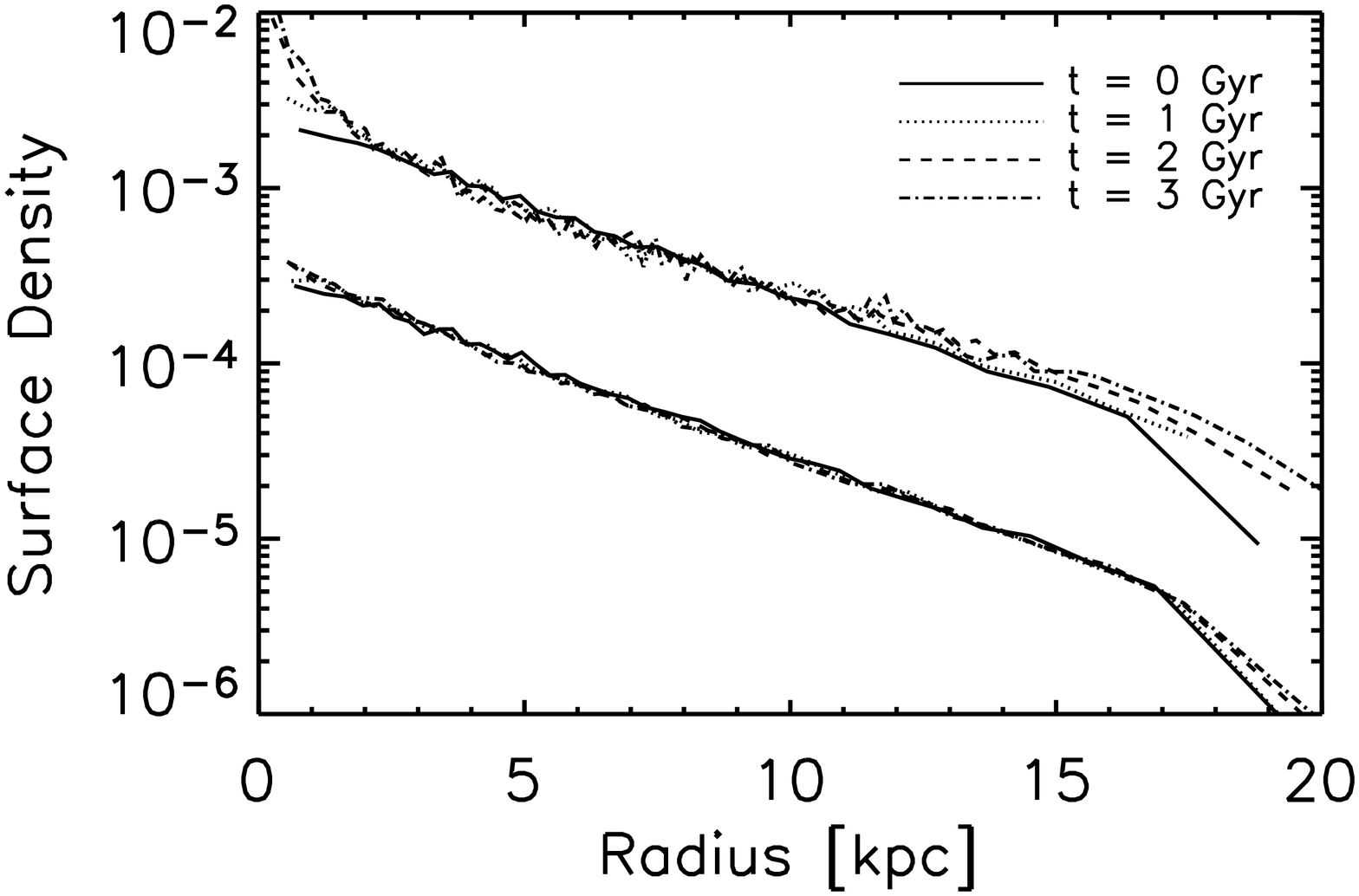} &&
\psfig{width=8cm,angle=90,file=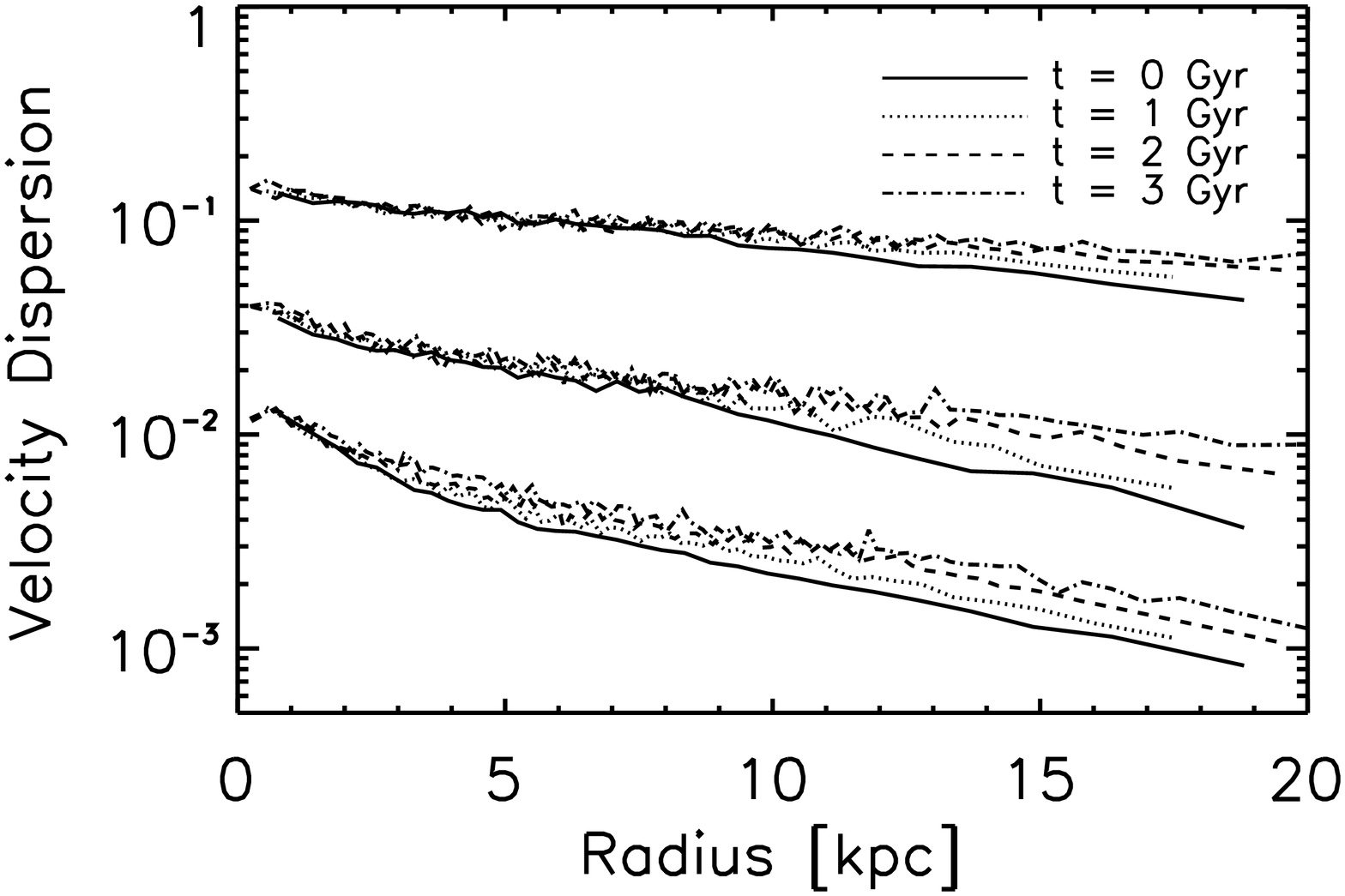} \\
\end{tabular}
\caption{ In the left plot the surface density of the stellar disk is
plotted for different times of $3$\,Gyr of evolution. The upper group
of curves is for the presented model including SF and all other
processes. For a comparison the lower group of curves (scaled by a
factor of $0.1$ to avoid overplotting) is for the pure N-body
evolution of the same galaxy model. The full model shows an increased
surface density in the center as well as in the outer parts. The right
diagram shows from top to bottom the velocity dispersions $\sigma_{\rm
R}$, $\sigma_{\rm \phi}$ (scaled with $0.2$) and $\sigma_{\rm z}$
(scaled with $0.1$). A heating of the disk can be seen.
}
\label{harfst_fig4}
\end{center}
\end{figure*}

The disk is stable over the full 3\,Gyr of evolution. \revised{This
can be seen in Fig.~\ref{harfst_fig4}, where a comparison of the
stellar surface density for the full model (i.e.\ including SF and all
other processes) with a pure N-body model is shown (left panel). In
the full model the surface density in the center (${\rm R} < 2\,{\rm
kpc}$) has increased significantly. The explanation for this is that
due to dissipation clouds have fallen towards the center before
forming stars. The surface density profiles of the clouds confirm this
idea. The velocity dispersion profiles (Fig.~\ref{harfst_fig4}, right
panel) show that the disk is heating up a little. A comparison
with the pure N-body model shows that some of the heating is due to
two-body relaxation but a significant amount is also due to the
kinetic feedback of SF. Simulations with higher initial partical
numbers should be performed to minimize effects of two-body relaxation
in order to determine the heating rate due to feedback.
}

\section{Summary and future Prospects}
\label{harfst_sec_sum}

We presented a particle code to model the evolution of galaxies
with a multi-phase description of the ISM including a new approach
to SF. A model of a Milky Way type galaxy shows a reasonable
behavior in terms of SF rate and stability of the disk.

In the near future more simulations are needed to determine how
the results depend on initial conditions, \revised{e.g. the
initial set-up for the hot gas or initial particle numbers} and
how other parameters in the model, e.g. the energy feedback by
SNe, affect the evolution of the model galaxy. This should result
in a reference model, that could be used in simulations of
interacting galaxies, e.g.\ to shed more light on trigger
mechanisms of star bursts.


\section*{Acknowledgments}

The authors would like to thank the organisers of the GCD-V meeting
for the opportunity to present our work and for conducting a
successful conference. S.~Harfst is grateful for financial support by
the {\it Deutsche Forsch\-ungs\-gemein\-schaft (DFG)} under grant
TH~511/2--4. S.~Harfst acknowledges the financial support of the
Australian Research Council through its Linkage International Award
scheme. We thank W.~Dehnen for making his TREE-code {\tt falcON}
publicly available.

\section*{References}






\reference Berczik, P., 1999, A\&A, 348, 371
\reference Berczik, P., Hensler, G., Theis, Ch. \& Spurzem, R., 2003, Ap\&SS, 284, 865
\reference B\"ohringer, H. \& Hensler, G., 1989, A\&A, 215, 147
\reference Cole, S., Aragon-Salamanca, A., Frenk, C.~S., Navarro, J.~F. \&
Zepf, S.~E., 1994, MNRAS, 271, 781
\reference Cowie, L.~L., McKee, C.~F. \& Ostriker, J.~P., 1981, ApJ, 247, 908
\reference Dehnen, W., 2002, JCP, 179, 27
\reference Elmegreen, B.~G. \& Efremov, Y.~N., 1997, ApJ, 480, 235
\reference Elmegreen, B.~G., 2000, ApJ, 530, 277
\reference Friedli, D. \& Benz, W., 1995, A\&A, 301, 649
\reference Gingold, R.~A. \& Monaghan, J.~J., 1977, MNRAS, 181, 375
\reference Hernquist, L. \& {Katz}, N., 1989, ApJSS, 70, 419
\reference Hultman, J. \& Pharasyn, A., 1999, A\&A, 347, 769
\reference Kroupa, P., Tout, C.~A. \& Gilmore, G., 1993, MNRAS, 262, 545
\reference Kuijken, K. \& Dubinski, J., 1995, MNRAS, 277, 1341
\reference Lucy, L.~B., 1977, AJ, 82, 1013
\reference Monaghan, J.~J., 1992, ARA\&A, 30, 543
\reference Navarro, J.~F., Frenk, C.~S. \& White, S.~D.~M., 1997,
ApJ, 490, 493
\reference Raiteri, C.~M., Villata, M. \& Navarro, J.~F., 1996, A\&A, 315, 105
\reference Ritchie, B.~W. \& Thomas, P.~A., 2001, MNRAS, 323, 743
\reference Rivolo, A.~R. \& Solomon, P.~M., 1988, in {\em Molecular Clouds in the Milky Way and External Galaxies}, eds. R.~L. Dickman, R.~L. Snell \& J.~S.
Young, p.~42
\reference Semelin, B. \& Combes, F., 2002, A\&A, 388, 826
\reference Springel, V. \& Hernquist, L., 2003, MNRAS, 339, 289
\reference Steinmetz, M. \& M\"uller, E., 1994, A\&A, 281, 97
\reference Steinmetz, M. \& M\"uller, E., 1995, MNRAS, 276, 549
\reference Theis, Ch. \& Hensler, G., 1993, A\&A, 280, 85
\reference White, S.~D.~M. \& Frenk, C.~S., 1991, ApJ, 379, 52

\end{document}